\documentclass[twocolumn,prl,aps,longbibliography,superscriptaddress]{revtex4-1}

\usepackage{times}

\usepackage{graphicx}
\usepackage{hyperref}
\usepackage{physics}

\usepackage{amsmath}
\usepackage{amsfonts}
\usepackage{amssymb}
\usepackage{amstext}
\usepackage{amsxtra}

\usepackage{mathtools}
\usepackage{isomath}
\usepackage{nccmath}
\usepackage{xspace}

\usepackage{wasysym}
\usepackage{xfrac}

\usepackage{float}
\usepackage{upgreek}

\usepackage{bm}
\usepackage{color}
 \usepackage{siunitx}

\hypersetup{backref,colorlinks=true,linkcolor=cyan,linktoc=black,citecolor=cyan,urlcolor=cyan}

\newcommand{\kB}{k_{\textrm{B}}}

\newcommand{\nth}{n_{\rm th}}%
\newcommand{\Nth}{N_{\rm th}}%
\newcommand{\Gth}{\Gamma_{\rm th}}%

\newcommand{\alg}[1]{\begin{align} #1 \end{align}}

\begin{document}
\setlength{\abovedisplayskip}{9pt + 0.6pt -4pt}
\setlength{\belowdisplayskip}{9pt + 0.6pt -4pt}

\title{Can three-body recombination purify a quantum gas?}

\author{Lena~H.~Dogra}
\affiliation{Cavendish Laboratory, University of Cambridge, J. J. Thomson Avenue, Cambridge CB3 0HE, United Kingdom}
\author{Jake~A.~P.~Glidden}
\affiliation{Cavendish Laboratory, University of Cambridge, J. J. Thomson Avenue, Cambridge CB3 0HE, United Kingdom}
\author{Timon~A.~Hilker}
\affiliation{Cavendish Laboratory, University of Cambridge, J. J. Thomson Avenue, Cambridge CB3 0HE, United Kingdom}
\author{Christoph~Eigen}
\affiliation{Cavendish Laboratory, University of Cambridge, J. J. Thomson Avenue, Cambridge CB3 0HE, United Kingdom}
\author{Eric~A.~Cornell}
\affiliation{JILA, National Institute of Standards and Technology and University of Colorado, and Department of Physics, Boulder, Colorado 80309-0440, USA}
\author{Robert~P.~Smith}
\affiliation{Clarendon Laboratory, University of Oxford, Parks Road, Oxford OX1 3PU, United Kingdom}
\author{Zoran~Hadzibabic}
\affiliation{Cavendish Laboratory, University of Cambridge, J. J. Thomson Avenue, Cambridge CB3 0HE, United Kingdom}

\begin{abstract}
Three-body recombination in quantum gases is traditionally associated with heating, but it was recently found that it can also cool the gas. We show that in a partially condensed three-dimensional homogeneous Bose gas three-body loss could even purify the sample, that is, reduce the entropy per particle and increase the condensed fraction $\eta$.
We predict that the evolution of  $\eta$ under continuous three-body loss can, depending on small changes in the initial conditions, exhibit two qualitatively different behaviours - if it is initially above a certain critical value, $\eta$ increases further, whereas clouds with lower initial $\eta$ evolve towards a thermal gas.
These dynamical effects should be observable under realistic experimental conditions.
\end{abstract}

\maketitle

In ultracold atomic gases, uncontrollable particle loss is usually associated with mundane and adverse effects, such as increase of temperature and entropy per particle.
However, it can also have more interesting consequences.
In a 3D weakly interacting homogeneous Bose gas, one-body loss due to collisions with the background gas in the vacuum chamber results in the quantum analogue of Joule-Thomson cooling ~\cite{Kothari:1937,Schmidutz:2014}. This is a purely quantum-statistical effect, with the only role of weak interactions being to ensure thermalisation of the gas. 
Recently, it was also observed that in interaction-dominated 1D Bose gases atom loss led to cooling even though its origin was three-body recombination, which is traditionally associated with heating~\cite{Schemmer:2018}.
In these experiments~\cite{Schmidutz:2014,Schemmer:2018}, losses reduced the gas temperature, but they still made the samples {\it less degenerate}, because the fractional drop of the degeneracy temperature, set by the gas density, was even larger.

\begin{figure}[b!]
\centering
\includegraphics[width=\columnwidth]{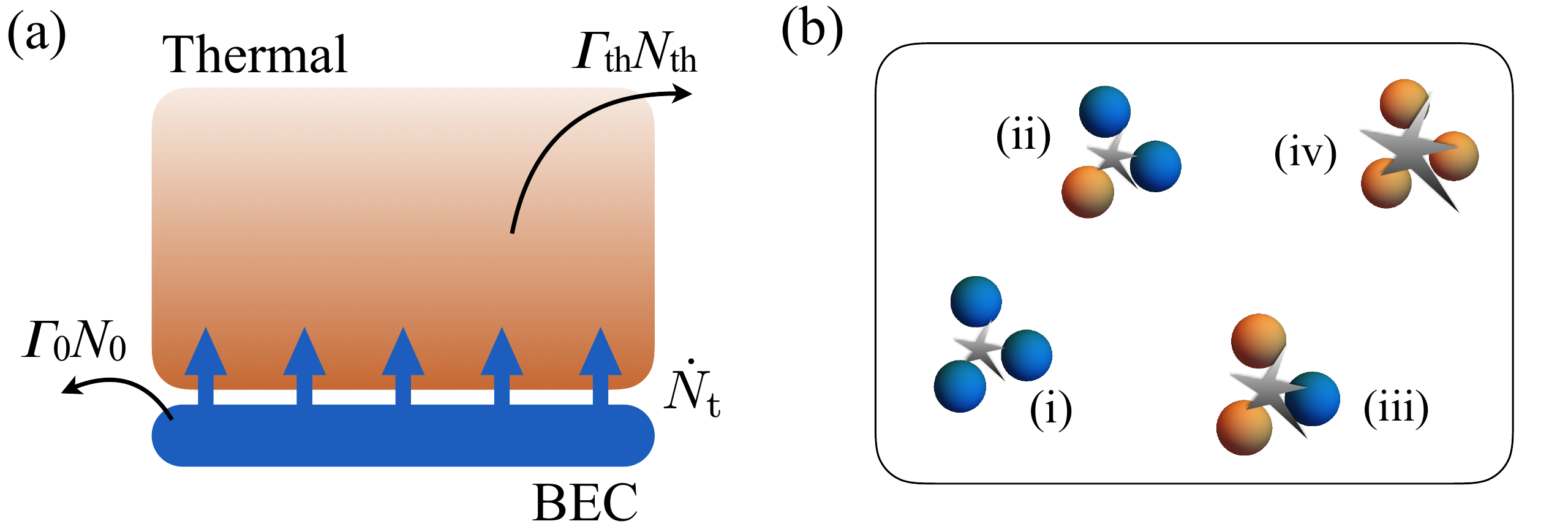}
\caption{Microscopic dynamics of an ideal homogeneous Bose gas with three-body loss. 
(a) Saturation-driven cooling. Loss of atoms from a saturated thermal cloud (at a rate $\Gth\Nth$) induces a flow of zero-energy atoms ($\dot{N}_{\rm t}$) from the BEC to the thermal gas, which lowers the temperature. Direct loss of the BEC atoms ($\Gamma_0 N_0$) has no effect on the temperature. 
(b) Three-body loss processes. The rates of three-body collisions between different numbers of BEC (blue) and thermal (orange) atoms involve different combinatorial terms, reflecting the boson bunching that occurs in a thermal cloud but not in a BEC. Normalised by the appropriate powers of BEC and thermal densities, the relative rates of the processes (i) - (iv) are, respectively,  1/(3!), 1/(2!), 1, and 1. This preferential loss of thermal atoms can lead to purification of the gas.}
\label{fig:1}
\end{figure}

In this Letter, we show that in a partially condensed, weakly interacting homogeneous 3D Bose gas, three-body recombination 
can result in an intricate dynamical phase diagram; under certain conditions it can both cool and purify the gas, {\it i.e.}  reduce the entropy per particle and increase the condensed fraction $\eta$.
An ideal-gas thermodynamic calculation gives that the evolution of the system depends on whether $\eta$ is above or below a critical value $\eta^* = 0.76$. For $\eta < \eta^*$, the gas cools but $\eta \rightarrow 0$. However, for $\eta > \eta^*$ the gas undergoes self-purification and $\eta \rightarrow 1$.
This behaviour is a consequence of the interplay of two quantum-statistical effects -- saturation of the thermal cloud~\cite{Pethick:2002, Schmidutz:2014} and preferential loss of thermal atoms due to boson bunching~\cite{Kagan:1985,Burt:1997,Soding:1999,Haller:2011} (see Fig.~\ref{fig:1}).
Purification occurs not just despite the three-body nature of the loss, but specifically because of it.  
Considering the effects of weak two-body interactions on the thermodynamics, we find a more complex phase diagram, but qualitatively similar behaviour for $na^3 < 10^{-7}$, where $n$ is the gas density and $a$ the $s$-wave scattering length.

These effects could be observed in a homogeneous Bose gas, produced in an optical box trap~\cite{Gaunt:2013}, near a zero-crossing of $a$ associated with a Feshbach resonance~\cite{Chin:2010}. For both the saturation of the thermal component and the beneficial effects of boson bunching for purification, it is important that the gas is homogeneous, with the condensed and thermal components completely spatially overlapped~\footnote{Due to geometric effects, in a harmonic trap the thermal atom number $\Nth$ is not saturated even for very weak interactions~\cite{Tammuz:2011}, whereas in a box-trapped gas it is~\cite{Schmidutz:2014}.}.  The gas homogeneity also eliminates the problem of `anti-evaporation' heating present in harmonic traps~\cite{Weber:2003}, where the density dependent recombination preferentially removes atoms with below-average energy. We assume that three-body recombination is the dominant loss process and that loss products leave the box without undergoing secondary collisions. At the end of the paper we discuss how these requirements can be fulfilled.

To elucidate the key physics, we start with an ideal-gas calculation, assuming that continuous thermalisation is the only effect of two-body interactions.

In Fig.~\ref{fig:1}(a) we outline the idea of saturation-driven cooling. 
In a partially condensed ideal Bose gas of $N$ atoms at temperature $T$, the thermal atom number $\Nth$ is saturated at the critical value for condensation $N_{\rm c}(T) \propto T^{\alpha}$, with $\alpha =3/2$, and there are $N_0 = N - N_{\rm c}$ zero-energy atoms in the Bose-Einstein condensate (BEC).
The total energy is $E \propto \Nth T \propto T^{\alpha+1}$ and the entropy per particle is proportional to the thermal fraction $1 - \eta = \Nth/N$~\cite{Pethick:2002}.
Removing BEC atoms through some loss process, at a rate we write as $\Gamma_0 N_0$, although $\Gamma_0$ may not be a constant, does not change $E$, $\Nth$ or $T$. However, removing thermal atoms through some (same or different) loss process, at a rate $\Gth \Nth$, reduces the energy according to $\dot{E}/E = -\Gth$.
Since $T \propto E^{1/(\alpha +1)}$ and $\Nth \propto E^{\alpha/(\alpha + 1)}$ depend only on $E$, we get
\begin{equation}
\frac{\dot{T}}{T}  = - \frac{1}{\alpha+1} \, \Gth <0 
\quad {\rm and} \quad \frac{\dot{N}_{\rm th}}{\Nth} = - \frac{\alpha}{\alpha+1} \Gth . 
\label{eq_JT}
\end{equation}
Note that $\dot{N}_{\rm th}/\Nth = -(3/5) \Gth \neq - \Gth$. To maintain equilibrium, with $\Nth$ saturated, atoms transfer between the BEC and the thermal cloud, at a rate $\dot{N}_{\rm t}$, so the net rates of change of $N_0$ and $\Nth$ are 
$\dot{N}_0 = - \Gamma_0 N_0 -\dot{N}_{\rm t}$ and $\dot{N}_{\rm th} =  -\Gth \Nth + \dot{N}_{\rm t}$.
 Specifically, for every 5 atoms lost from the thermal cloud, 2 are replenished from the BEC. This injection of zero-energy particles into the thermal cloud is the microscopic origin of the cooling. 

These arguments are not specific to any particular loss process. They apply to the three-body loss discussed here and the one-body loss that drives the quantum Joule-Thomson effect observed in~\cite{Schmidutz:2014}, and are also at the heart of the decoherence-driven cooling observed in~\cite{Lewandowski:2003,Olf:2015}, although in that case the atoms were not lost, but transferred to a different spin state.

To see whether atom loss can purify the gas, we calculate
\alg{
	\frac{\dot{\bar{\eta}}}{\bar{\eta}} =  \frac{\dot{N}_{\rm th}}{\Nth} - \frac{\dot{N}}{N}   
= \Gamma (1-\mathcal{P}) \, , 
\label{eq_eta}
}
where $\bar{\eta} = 1 -\eta$ is the thermal fraction, $\Gamma = - \dot{N}/N =  (N_0\Gamma_0 + \Nth \Gth)/N$ is the total per-particle loss rate, and we have introduced a dimensionless purification coefficient
\alg{
\mathcal{P} \equiv \frac{\dot{N}_{\rm th}/\Nth}{\dot{N}/N} \, , \quad {\rm so} 
\quad \mathcal{P} - 1= \frac{{\mathrm d}[\ln(1 - \eta)]}{{\mathrm d}[\ln(N)]} \, .
 \label{eq_betadef}
}
For $\mathcal{P} > 1$ the gas purifies ($\dot{\eta} > 0$), whereas for $0 < \mathcal{P} < 1$ it cools without purifying. From Eq.~(\ref{eq_JT}), for an ideal gas
\alg{
\mathcal{P} = \frac{\alpha}{\alpha+1}\frac{\Gth}{\Gamma} = \frac35 \, \frac{\Gth}{\Gamma} \, ,
 \label{eq_beta}
}
so purification requires $\Gth/\Gamma > 5/3$.
Here the nature of the loss process is crucial. 
One-body losses do not distinguish BEC and thermal atoms, so $\Gth = \Gamma_0 = \Gamma$ and $\mathcal{P} = 3/5$. However, for three-body loss $\mathcal{P}$ can be larger than $1$.

In general, the local three-body loss rate is given by 
 \begin{equation}
 	\dot{n}/n =  -  g_3 K_3 n^2 \label{eq_3bl} \, 
\end{equation} 
where $g_3$ is the zero-distance three-body correlation function and $K_3$ is the three-body loss coefficient. In terms of local condensate and thermal density, $n_0$ and $\nth$ respectively~\cite{Kagan:1985},
\alg{ 
g_3 = \frac{3!}{n^3} \qty(\frac1{3!}n_0^3  +  \frac1{2!} 3 n_0^2 \nth + 3 n_0\nth^2 + \nth^3) \, . \label{eq_g3}
}
For a uniform gas, where $n_0/N_0 = \nth/\Nth = n/N = 1/V$, with $V$ being the gas volume, this corresponds to 
\begin{equation}
\Gamma = K_3 n^2\qty(6-9\eta^2+4\eta^3)  \, . \label{eq_Gamma}
\end{equation}
For the same $N$ and $V$, the loss rate in a pure BEC ($\eta=1$) is 6 times smaller than in a thermal gas ($\eta=0$), due to suppression of boson bunching~\cite{Kagan:1985,Burt:1997}. 
More generally, the four terms on the r.h.s.~of Eq.~(\ref{eq_g3}) correspond, left to right, to the four loss processes (i) - (iv) in Fig.~\ref{fig:1}(b).
Considering how many thermal and BEC atoms are lost in each process and keeping the same order of terms as in Eq.~(\ref{eq_g3}):
\begin{align}
\Gamma_0 N_0 &= K_3 \left( N_0^3 + 6 N_0^2 \Nth + 6 N_0 N_{\rm th}^2  +0 \right)/V^2  \, ,\nonumber \\
\Gth \Nth &= K_3 \left( 0 +  3 N_0^2 \Nth + 12 N_0 N_{\rm th}^2  + 6 N_{\rm th}^3\right )/V^2 \, ,\nonumber 
\end{align}
corresponding to
\alg{
	\Gamma_0 &= K_3n^2\qty(6-6\eta+\eta^2) \, , \nonumber\\
	\Gth &= K_3n^2\qty(6-3\eta^2) \label{eq_LossRates}  \, .
}
Finally, inserting $\Gamma$ and $\Gth$ into Eqs.~(\ref{eq_JT}, \ref{eq_eta}, \ref{eq_beta}), we obtain: 
\alg{
	\frac{\dot{T}}{T} &= - K_3{n^2}\frac{6}{5}(2-\eta^2)\label{eq_res}  \, ,  \nonumber  \\ 
	\frac{\dot{\bar{\eta}}}{\bar{\eta}} &= -K_3{n^2}\frac45(-3+9\eta^2-5\eta^3)  \, , \\ %
	\mathcal{P} &= \frac{9}{5} \, \frac{2 - \eta^2}{6 - 9\eta^2 + 4\eta^3} \;  \nonumber. 
}

\begin{figure}
\includegraphics[width=\columnwidth]{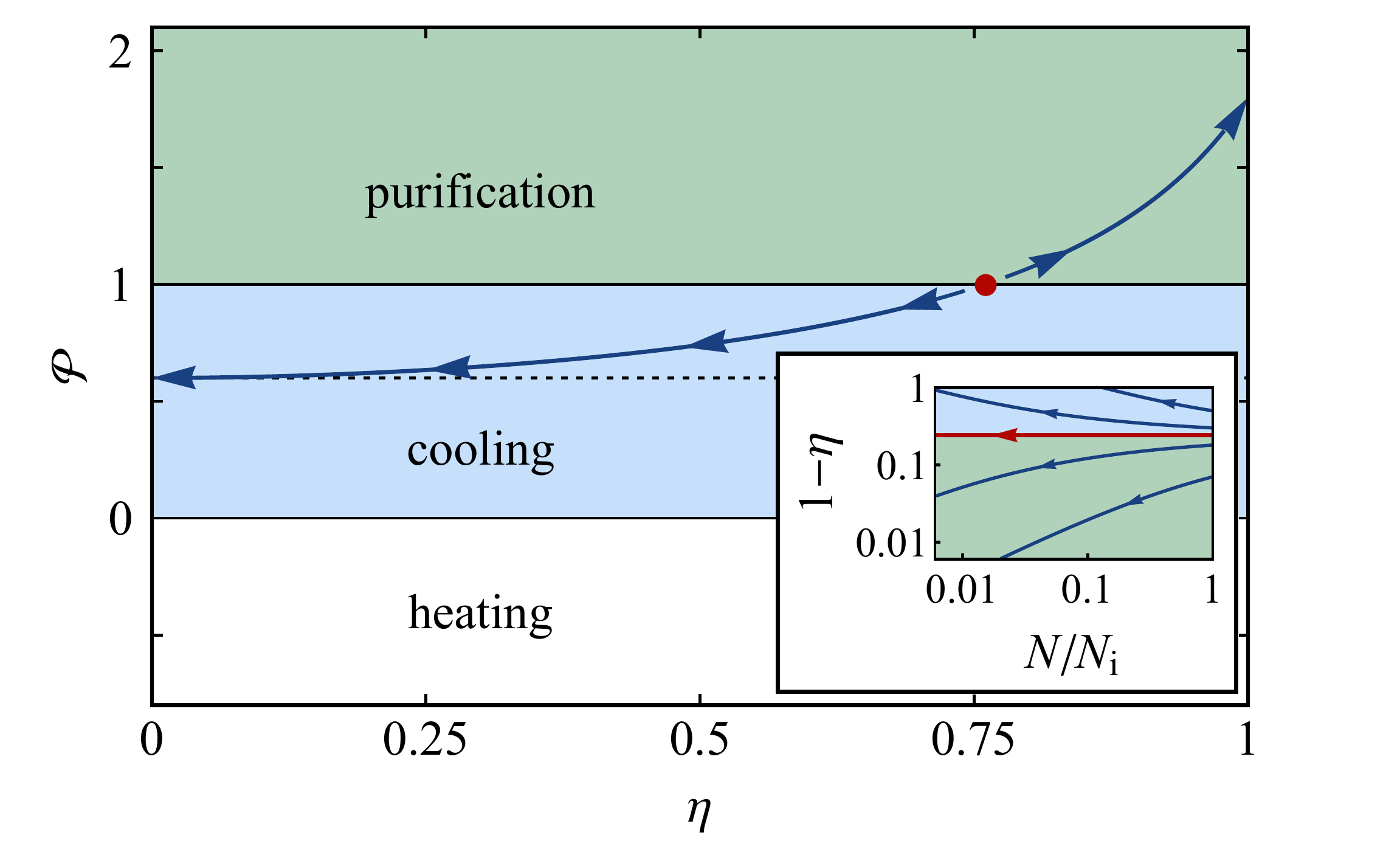}
\caption{
Three-body cooling and purification of an ideal partially condensed homogeneous Bose gas.
The purification coefficient $\mathcal{P}$ (see text), which determines whether the gas purifies ($\mathcal{P}>1$) or cools without purifying ($0<\mathcal{P} <1$), depends only on the condensed fraction $\eta$. The critical value $\eta^* =0.76$ (indicated by the red dot) defines a `bifurcation point' for the evolution of the cloud. As indicated by the arrows, for $\eta < \eta^*$ the condensed fraction keeps dropping, but for $\eta > \eta^*$ the gas self-purifies and $\eta \rightarrow 1$. The horizontal dashed line, $\mathcal{P} = 3/5$, corresponds to the result for one-body loss, which cannot purify the gas. Inset: evolution of $\eta$ for different initial conditions; here $N_{\rm i}$ is the initial atom number. 
}
\label{fig:2}
\end{figure}

We see that $\mathcal{P}$ depends only on the condensed fraction $\eta$. As shown in Fig.~\ref{fig:2}, it monotonically grows from $3/5$ at $\eta =0$ to $9/5$ at $\eta=1$~\footnote{One can repeat an analogous calculation for two-body losses due to, {\it e.g.}, spin-changing collisions. In that case $\dot{n}/n = -g_2 K_2 n$, with $g_2 = (n_0^2 + 4 n_0 n_{\rm th} + 2 n_{\rm th}^2)/n^2 = 2-\eta^2$. This gives $\mathcal{P} = 6/(10-5\eta^2)$, which can also be larger than 1.}.
For very small $\eta$, from $N \approx \Nth$ it directly follows that $\Gamma \approx \Gth$ and $\mathcal{P} \approx 3/5$. In this regime also $\Gamma_0 \approx \Gth \approx 6K_3 \nth^2$. Microscopically,  in this regime the two dominant processes in Fig.~\ref{fig:1}(b) are (iii) for the loss of BEC atoms and (iv) for the loss of thermal ones. These involve at most one BEC atom and hence have the same combinatorial factors, so $\Gamma_0 \approx \Gth$, and we essentially get the quantum Joule-Thomson effect~\cite{Schmidutz:2014}, although driven by three-body loss. In the opposite limit $\eta \approx 1$, where $N \approx N_0$ and $\Gamma \approx \Gamma_0$, the two relevant processes in Fig.~\ref{fig:1}(b) are (i) and (ii), which have different combinatorial factors, such that $\Gth \approx 3 \Gamma_0 \approx 3 \Gamma$, giving $\mathcal{P} \approx 9/5$.

Crucially, $\mathcal{P} - 1$ changes sign at a critical condensed fraction $\eta^* = 0.76$, which is a solution to the cubic equation $\dot{\bar{\eta}} = 0$, see Eq.~(\ref{eq_res}). As indicated by the arrows in Fig.~\ref{fig:2}, for $\eta < \eta^*$ the gas cools but $\eta \rightarrow 0$, while for $\eta > \eta^*$ the gas keeps self-purifying and $\eta \rightarrow 1$. This is illustrated in the inset of Fig.~\ref{fig:2}, where we show the evolution of the thermal fraction for different initial condensed fractions. On this log-log plot, $\mathcal{P} -1$
gives the slope of the $\bar{\eta}(N)$ trajectories; see Eq.~(\ref{eq_betadef}).

These ideal-gas effects should play a dominant role if the interaction energy is small compared to the thermal one. Within mean-field theory (see below), for small thermal fraction the ratio  of thermal to interaction energy is $\approx\!0.4~\bar{\eta}^{5/3} /(na^3)^{1/3}$~\cite{Pethick:2002}, so the two are comparable for $\bar{\eta} = (na^3)^{1/5}$.

\begin{figure}[t!]
\centering
\includegraphics[width=\columnwidth]{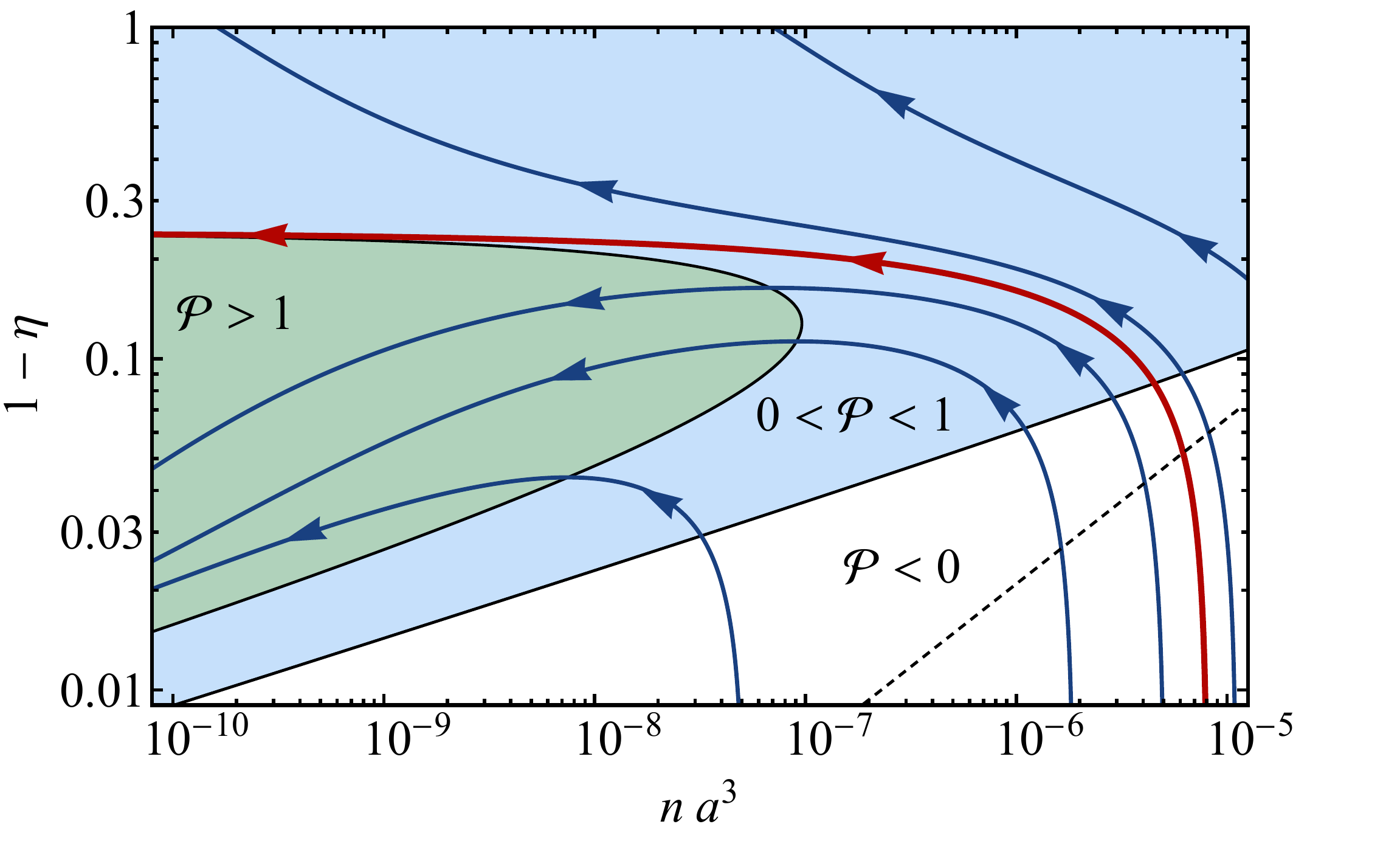}
\caption{
Cooling and purification in a weakly interacting gas. The purification coefficient $\mathcal{P}$ now depends on the condensed fraction $\eta$ and the gas parameter $na^3$. The evolution of the system is described by trajectories that flow either to $\eta =1$ or to $\eta=0$, depending on which side of the critical trajectory $\eta^*(na^3)$ (red line) they are. On this log-log graph the slope of the trajectories is given by $\mathcal{P}-1$ and the background shading indicates whether instantaneously the gas purifies (green), cools without purifying (blue), or heats (white). 
In the regime below the dashed line, which corresponds to $\kB T = 2gn$, the dynamics could deviate from our results due to phononic nature of the thermal excitations.
}
\label{fig:3}
\end{figure}

We now quantitatively assess the effects of weak two-body interactions on three-body cooling and purification, for $na^3\lesssim 10^{-5}$ (see Fig.~\ref{fig:3}). In this regime, to a good approximation, interaction energy is mean-field like, $g_3$ is ideal-gas like~\cite{Kagan:1985, Haller:2011}, and the saturation picture holds~\cite{Smith:2017}. We also assume that the thermal excitations are particle-like, which is a good approximation for most of the range of system parameters we consider (see dashed line in Fig.~\ref{fig:3}).
The total energy is now
\alg{
E & = \alpha_0 \Nth  \kB T  + \frac{g}{2 V} \left( N_0^2 + 4 N_0 \Nth + 2 \Nth^2 \right) \, . \label{eq_etot1}
}
Here $\alpha_0 = \alpha \zeta(\alpha + 1)/\zeta(\alpha) = 0.77$, where $\zeta$ is the Riemann function, and $g = 4 \pi \hbar^2 a/m$, where $m$ is the atom mass.

A subtle question is how much interaction energy is removed from the gas through atom loss. Let us first consider an initially pure BEC, with $E = g N_0^2/(2V)$. For the BEC to stay pure after removal of a particle, the energy removed would have to be $\mu = \partial E/\partial N_0$. This would correspond to removing a particle adiabatically from a delocalised wavefunction. In contrast, a sudden local atom loss should simply remove the average energy per particle, $E/N_0= \mu/2$. The gas is then left with total energy larger, by $\mu/2$, than that of a pure BEC with $ N_0 - 1$ atoms, so this loss leads to heating.
The next conceptual step is to extend this analysis to nonzero $T$. We rewrite Eq.~(\ref{eq_etot1}) as
\alg{
E \! = \! \left[ \frac{g}{V} \! \left( \frac12 N_0 \! + \! \Nth \right) \! \right] \! N_0 + \left[ \vphantom{\frac12}  \alpha_0 \kB  T + \frac{g}{V} \! \left(N_0 \! + \! \Nth \right) \! \right] \! \Nth \nonumber} 
and interpret the terms in square brackets as the energy per BEC atom, $\varepsilon_0$ (left bracket), and the energy per thermal atom,  $\varepsilon_{\rm th}$ (right bracket), in the sense that the rate of energy change should be
\alg{
 \dot{E} = -  \varepsilon_0 \Gamma_0 N_0  -  \varepsilon_{\rm th} \Gth \Nth \, .\label{eq_dEdt}
}
Under continuous equilibration it must also be
\alg{
\dot{E} \! = \! \frac{\partial E}{\partial N_0} \!\left(-\Gamma_0 N_0 - \dot{N}_{\rm t} \right) \!+ \frac{\partial E}{\partial \Nth}\!\left(- \Gth \Nth + \dot{N}_{\rm t}\right) \!, 
}
where $\dot{N}_{\rm t}$ is such that $\Nth$ remains saturated, and it can now in general be of either sign.
Combining these equations gives the purification coefficient $\mathcal{P}$,  a generalisation of Eq.~(\ref{eq_res}), which now depends on two dimensionless parameters, $\eta$ and $na^3$:
\alg{
 \mathcal{P} = \frac{  9  \left(2 - \eta ^2\right)   + \, b_1(\eta) \, (n a^3)^{1/3}}{  5  \left( 6 - 9\eta^2 + 4 \eta ^3 \right) +  \, b_2(\eta) \, (n a^3)^{1/3}} , \label{eq_betaint} 
}
where $b_1  (\eta) \! = \!  \gamma \left( 7 \eta ^4  \! - \!  20 \eta ^3   \! +  \! 12 \eta ^2  \! + \!  12 \eta\!   -\!   12 \right)\left(1-\eta \right)^{- 5/3} $ and $b_2 (\eta) \! =\!  2 \gamma \eta (6 -9 \eta^2  + 4 \eta^3)\left(1-\eta \right)^{-2/3}$, with $\! \gamma \! = \! 2 \zeta(3/2)^{5/3} / \zeta(5/2) = 7.4$.

In Fig.~\ref{fig:3} we show examples of trajectories $\eta(na^3)$ for fixed (arbitrary) $a$. The red-coloured trajectory separates those that flow to $\eta =0$ and $\eta =1$. The background shading indicates whether the gas instantaneously purifies ($\mathcal{P} >1$), cools but does not purify ($0< \mathcal{P} < 1$), or heats ($\mathcal{P} <0$)~\footnote{One can obtain identical results by iteratively removing, in small steps, particles according to Eq.~(\ref{eq_LossRates}) and energy according to Eq.~(\ref{eq_dEdt}), and then solving for the new equilibrium state under the constraints of the new total $N$ and $E$.}. 

At low thermal fraction $\bar{\eta}$, the constant-$\mathcal{P}$ contours in Fig.~\ref{fig:3} follow the scaling $\bar{\eta} \propto (na^3)^{1/5}$, meaning that $\mathcal{P}$ is determined by the ratio of thermal and interaction energies. Qualitatively, affinity between particles (due to quantum statistics) leads to cooling, while aversion (due to repulsive interactions) leads to heating, similarly to how Joule-Thomson rarefaction leads to cooling of attractive classical gases and non-interacting bosons, and heating of repulsive classical gasses and non-interacting fermions~\cite{Kothari:1937, Schmidutz:2014}; here, each of the two opposing effects dominates in a different regime. The $\mathcal{P}=0$ contour is $\bar{\eta} \approx (na^3)^{1/5}$ all the way to $na^3 = 10^{-5}$, while the purification effect is less robust in presence of two-body repulsion, but is still possible for $na^3 < 10^{-7}$. Also note that a system trajectory cannot leave the purification region $\mathcal{P}>1$, but can enter it because losses reduce $na^3$.  We have considered particle-like excitations, while phononic excitations will dominate the system's evolution for small $T /(gn) \sim (\bar{\eta}/\sqrt{na^3})^{2/3}$, below the dashed line in Fig.~\ref{fig:3}.

Our theory could be tested near a zero-crossing of $a$, associated with a Feshbach resonance, where $K_3$ is nonzero and nearly $a$-independent. For illustration, we assume $K_3 \approx \SI{d-29}{\centi \meter^6\per \second}$, as observed in, {\it e.g.}, $^7{\rm Li}$~\cite{Shotan:2014} and  $^{39}{\rm K}$~\cite{Fattori:2008}, initial $n = \SI{e14}{\per\cubic \centi \metre}$ and $\eta = 0.9$, and $a=10\,a_0$, where $a_0$ is the Bohr radius. 
For these parameters, $na^3 =  \num{1.5d-8}$, our calculation gives $\mathcal{P} > 1$ (see Fig.~\ref{fig:3}), and $\Gamma \approx \SI{0.1}{\per \second}$ would be sufficiently large to dominate over the one-body loss rate, which is in many experiments $< \SI{0.01}{\per \second}$.
The healing length would be $\xi = 1/\sqrt{8 \pi n_0 a} \approx \SI{1}{\micro\metre}$, so in a box of size $L \gtrsim  \SI{10}{\micro\metre}$ the BEC would be essentially homogeneous and occupy the same volume as the thermal gas. The mean free path would be $\ell = 1/(8 \pi n a^2) \approx \SI{1}{\milli \metre}$, so secondary collisions of the loss products should be negligible. 
Finally, for continuous thermalisation  we want $\Gamma_2 > 3\, \dot{T}/T$~\cite{Monroe:1993}, where $\Gamma_2 \approx \bar{\eta} \sqrt{8 k_{\rm B} T /(\pi m)} 8 \pi n a^2$ (for small $\bar{\eta}$) is the per-particle rate of elastic two-body collisions, and $\dot{T}/T = \mathcal{P} \Gamma /\alpha \approx \Gamma$ from Eqs.~(\ref{eq_betadef}, \ref{eq_betaint}). This final requirement would be marginally satisfied in a $^{39}{\rm K}$ gas, and very comfortably in a $^7{\rm Li}$ one. We note that the initial $n$ we assume is a few times larger than what was already achieved in box traps, but is not unrealistic.


In conclusion, we have shown that,  under realistic experimental conditions, three-body recombination can both cool and purify a homogeneous Bose gas. We have calculated a dynamical phase diagram which shows that the behaviour of the system can be qualitatively altered by small changes in the initial conditions. An interesting extension of this work would be to investigate the regimes of stronger interactions and/or very low thermal fractions, where the phonon nature of the excitations plays a role, thus connecting our study with the analysis performed in Refs.~\cite{Schemmer:2018, Bouchoule:2018}.


We thank Jean Dalibard for helpful discussions. This work was supported by ERC (QBox), EPSRC [Grants No.~EP/N011759/1 and No.~EP/P009565/1], QuantERA [NAQUAS, EPSRC Grant No.~EP/R043396/1], AFOSR, and ARO. T.~A.~H. acknowledges support from the EU Marie Sk\l{}odowska-Curie program [Grant No.~MSCA-IF-2018 840081]. R.~P.~S. acknowledges support from the Royal Society. E.~A.~C. acknowledges hospitality and support from Trinity College, Cambridge. 
%


\vspace{-1.2em}
\begin{thebibliography}{22}%
\makeatletter
\providecommand \@ifxundefined [1]{%
 \@ifx{#1\undefined}
}%
\providecommand \@ifnum [1]{%
 \ifnum #1\expandafter \@firstoftwo
 \else \expandafter \@secondoftwo
 \fi
}%
\providecommand \@ifx [1]{%
 \ifx #1\expandafter \@firstoftwo
 \else \expandafter \@secondoftwo
 \fi
}%
\providecommand \natexlab [1]{#1}%
\providecommand \enquote  [1]{``#1''}%
\providecommand \bibnamefont  [1]{#1}%
\providecommand \bibfnamefont [1]{#1}%
\providecommand \citenamefont [1]{#1}%
\providecommand \href@noop [0]{\@secondoftwo}%
\providecommand \href [0]{\begingroup \@sanitize@url \@href}%
\providecommand \@href[1]{\@@startlink{#1}\@@href}%
\providecommand \@@href[1]{\endgroup#1\@@endlink}%
\providecommand \@sanitize@url [0]{\catcode `\\12\catcode `\$12\catcode
  `\&12\catcode `\#12\catcode `\^12\catcode `\_12\catcode `\%12\relax}%
\providecommand \@@startlink[1]{}%
\providecommand \@@endlink[0]{}%
\providecommand \url  [0]{\begingroup\@sanitize@url \@url }%
\providecommand \@url [1]{\endgroup\@href {#1}{\urlprefix }}%
\providecommand \urlprefix  [0]{URL }%
\providecommand \Eprint [0]{\href }%
\providecommand \doibase [0]{http://dx.doi.org/}%
\providecommand \selectlanguage [0]{\@gobble}%
\providecommand \bibinfo  [0]{\@secondoftwo}%
\providecommand \bibfield  [0]{\@secondoftwo}%
\providecommand \translation [1]{[#1]}%
\providecommand \BibitemOpen [0]{}%
\providecommand \bibitemStop [0]{}%
\providecommand \bibitemNoStop [0]{.\EOS\space}%
\providecommand \EOS [0]{\spacefactor3000\relax}%
\providecommand \BibitemShut  [1]{\csname bibitem#1\endcsname}%
\let\auto@bib@innerbib\@empty
\bibitem [{\citenamefont {Kothari}\ and\ \citenamefont
  {Srivasava}(1937)}]{Kothari:1937}%
  \BibitemOpen
  \bibfield  {author} {\bibinfo {author} {\bibfnamefont {D.~S.}\ \bibnamefont
  {Kothari}}\ and\ \bibinfo {author} {\bibfnamefont {B.~N.}\ \bibnamefont
  {Srivasava}},\ }\bibfield  {title} {\enquote {\bibinfo {title}
  {Joule--{T}homson {E}ffect and {Q}uantum {S}tatistics},}\ }\href
  {https://doi.org/10.1038/140970b0} {\bibfield  {journal} {\bibinfo  {journal}
  {Nature}\ }\textbf {\bibinfo {volume} {140}},\ \bibinfo {pages} {970}
  (\bibinfo {year} {1937})}\BibitemShut {NoStop}%
\bibitem [{\citenamefont {Schmidutz}\ \emph {et~al.}(2014)\citenamefont
  {Schmidutz}, \citenamefont {Gotlibovych}, \citenamefont {Gaunt},
  \citenamefont {Smith}, \citenamefont {Navon},\ and\ \citenamefont
  {Hadzibabic}}]{Schmidutz:2014}%
  \BibitemOpen
  \bibfield  {author} {\bibinfo {author} {\bibfnamefont {T.~F.}\ \bibnamefont
  {Schmidutz}}, \bibinfo {author} {\bibfnamefont {I.}~\bibnamefont
  {Gotlibovych}}, \bibinfo {author} {\bibfnamefont {A.~L.}\ \bibnamefont
  {Gaunt}}, \bibinfo {author} {\bibfnamefont {R.~P.}\ \bibnamefont {Smith}},
  \bibinfo {author} {\bibfnamefont {N.}~\bibnamefont {Navon}}, \ and\ \bibinfo
  {author} {\bibfnamefont {Z.}~\bibnamefont {Hadzibabic}},\ }\bibfield  {title}
  {\enquote {\bibinfo {title} {{Quantum Joule--Thomson Effect in a Saturated
  Homogeneous Bose Gas}},}\ }\href {\doibase 10.1103/PhysRevLett.112.040403}
  {\bibfield  {journal} {\bibinfo  {journal} {Phys. Rev. Lett.}\ }\textbf
  {\bibinfo {volume} {112}},\ \bibinfo {pages} {040403} (\bibinfo {year}
  {2014})}\BibitemShut {NoStop}%
\bibitem [{\citenamefont {Schemmer}\ and\ \citenamefont
  {Bouchoule}(2018)}]{Schemmer:2018}%
  \BibitemOpen
  \bibfield  {author} {\bibinfo {author} {\bibfnamefont {M.}~\bibnamefont
  {Schemmer}}\ and\ \bibinfo {author} {\bibfnamefont {I.}~\bibnamefont
  {Bouchoule}},\ }\bibfield  {title} {\enquote {\bibinfo {title} {{Cooling a
  Bose Gas by Three--Body Losses}},}\ }\href
  {https://doi.org/10.1103/PhysRevLett.121.200401} {\bibfield  {journal}
  {\bibinfo  {journal} {Phys. Rev. Lett.}\ }\textbf {\bibinfo {volume} {121}},\
  \bibinfo {pages} {200401} (\bibinfo {year} {2018})}\BibitemShut {NoStop}%
\bibitem [{\citenamefont {Pethick}\ and\ \citenamefont
  {Smith}(2002)}]{Pethick:2002}%
  \BibitemOpen
  \bibfield  {author} {\bibinfo {author} {\bibfnamefont {C.~J.}\ \bibnamefont
  {Pethick}}\ and\ \bibinfo {author} {\bibfnamefont {H.}~\bibnamefont
  {Smith}},\ }\href@noop {} {\emph {\bibinfo {title} {{B}ose--{E}instein
  Condensation in {D}ilute {G}ases}}}\ (\bibinfo  {publisher} {Cambridge
  University Press},\ \bibinfo {year} {2002})\BibitemShut {NoStop}%
\bibitem [{\citenamefont {Kagan}\ \emph {et~al.}(1985)\citenamefont {Kagan},
  \citenamefont {Svistunov},\ and\ \citenamefont {Shlyapnikov}}]{Kagan:1985}%
  \BibitemOpen
  \bibfield  {author} {\bibinfo {author} {\bibfnamefont {Y.}~\bibnamefont
  {Kagan}}, \bibinfo {author} {\bibfnamefont {B.~V.}\ \bibnamefont
  {Svistunov}}, \ and\ \bibinfo {author} {\bibfnamefont {G.~V.}\ \bibnamefont
  {Shlyapnikov}},\ }\bibfield  {title} {\enquote {\bibinfo {title} {Effect of
  {B}ose condensation on inelastic processes in gases},}\ }\href
  {http://www.jetpletters.ac.ru/ps/1420/article_21579.shtml} {\bibfield
  {journal} {\bibinfo  {journal} {JETP Lett.}\ }\textbf {\bibinfo {volume}
  {42}},\ \bibinfo {pages} {209} (\bibinfo {year} {1985})}\BibitemShut
  {NoStop}%
\bibitem [{\citenamefont {Burt}\ \emph {et~al.}(1997)\citenamefont {Burt},
  \citenamefont {Ghrist}, \citenamefont {Myatt}, \citenamefont {Holland},
  \citenamefont {Cornell},\ and\ \citenamefont {Wieman}}]{Burt:1997}%
  \BibitemOpen
  \bibfield  {author} {\bibinfo {author} {\bibfnamefont {E.~A.}\ \bibnamefont
  {Burt}}, \bibinfo {author} {\bibfnamefont {R.~W.}\ \bibnamefont {Ghrist}},
  \bibinfo {author} {\bibfnamefont {C.~J.}\ \bibnamefont {Myatt}}, \bibinfo
  {author} {\bibfnamefont {M.~J.}\ \bibnamefont {Holland}}, \bibinfo {author}
  {\bibfnamefont {E.~A.}\ \bibnamefont {Cornell}}, \ and\ \bibinfo {author}
  {\bibfnamefont {C.~E.}\ \bibnamefont {Wieman}},\ }\bibfield  {title}
  {\enquote {\bibinfo {title} {{Coherence, Correlations, and Collisions: What
  One Learns about Bose--Einstein Condensates from Their Decay}},}\ }\href
  {\doibase 10.1103/PhysRevLett.79.337} {\bibfield  {journal} {\bibinfo
  {journal} {Phys. Rev. Lett.}\ }\textbf {\bibinfo {volume} {79}},\ \bibinfo
  {pages} {337--340} (\bibinfo {year} {1997})}\BibitemShut {NoStop}%
\bibitem [{\citenamefont {S{{\"o}}ding}\ \emph {et~al.}(1999)\citenamefont
  {S{{\"o}}ding}, \citenamefont {Gu{\'e}ry-Odelin}, \citenamefont {Desbiolles},
  \citenamefont {Chevy}, \citenamefont {Inamori},\ and\ \citenamefont
  {Dalibard}}]{Soding:1999}%
  \BibitemOpen
  \bibfield  {author} {\bibinfo {author} {\bibfnamefont {J.}~\bibnamefont
  {S{{\"o}}ding}}, \bibinfo {author} {\bibfnamefont {D.}~\bibnamefont
  {Gu{\'e}ry-Odelin}}, \bibinfo {author} {\bibfnamefont {P.}~\bibnamefont
  {Desbiolles}}, \bibinfo {author} {\bibfnamefont {F.}~\bibnamefont {Chevy}},
  \bibinfo {author} {\bibfnamefont {H.}~\bibnamefont {Inamori}}, \ and\
  \bibinfo {author} {\bibfnamefont {J.}~\bibnamefont {Dalibard}},\ }\bibfield
  {title} {\enquote {\bibinfo {title} {Three-body decay of a rubidium
  {B}ose--{E}instein condensate},}\ }\href
  {https://doi.org/10.1007/s003400050805} {\bibfield  {journal} {\bibinfo
  {journal} {Appl. Phys. B}\ }\textbf {\bibinfo {volume} {69}},\ \bibinfo
  {pages} {257} (\bibinfo {year} {1999})}\BibitemShut {NoStop}%
\bibitem [{\citenamefont {Haller}\ \emph {et~al.}(2011)\citenamefont {Haller},
  \citenamefont {Rabie}, \citenamefont {Mark}, \citenamefont {Danzl},
  \citenamefont {Hart}, \citenamefont {Lauber}, \citenamefont {Pupillo},\ and\
  \citenamefont {N\"agerl}}]{Haller:2011}%
  \BibitemOpen
  \bibfield  {author} {\bibinfo {author} {\bibfnamefont {E.}~\bibnamefont
  {Haller}}, \bibinfo {author} {\bibfnamefont {M.}~\bibnamefont {Rabie}},
  \bibinfo {author} {\bibfnamefont {M.~J.}\ \bibnamefont {Mark}}, \bibinfo
  {author} {\bibfnamefont {J.~G.}\ \bibnamefont {Danzl}}, \bibinfo {author}
  {\bibfnamefont {R.}~\bibnamefont {Hart}}, \bibinfo {author} {\bibfnamefont
  {K.}~\bibnamefont {Lauber}}, \bibinfo {author} {\bibfnamefont
  {G.}~\bibnamefont {Pupillo}}, \ and\ \bibinfo {author} {\bibfnamefont
  {H.-C.}\ \bibnamefont {N\"agerl}},\ }\bibfield  {title} {\enquote {\bibinfo
  {title} {{Three--Body Correlation Functions and Recombination Rates for
  Bosons in Three Dimensions and One Dimension}},}\ }\href
  {https://link.aps.org/doi/10.1103/PhysRevLett.107.230404} {\bibfield
  {journal} {\bibinfo  {journal} {Phys. Rev. Lett.}\ }\textbf {\bibinfo
  {volume} {107}},\ \bibinfo {pages} {230404} (\bibinfo {year}
  {2011})}\BibitemShut {NoStop}%
\bibitem [{\citenamefont {Gaunt}\ \emph {et~al.}(2013)\citenamefont {Gaunt},
  \citenamefont {Schmidutz}, \citenamefont {Gotlibovych}, \citenamefont
  {Smith},\ and\ \citenamefont {Hadzibabic}}]{Gaunt:2013}%
  \BibitemOpen
  \bibfield  {author} {\bibinfo {author} {\bibfnamefont {A.~L.}\ \bibnamefont
  {Gaunt}}, \bibinfo {author} {\bibfnamefont {T.~F.}\ \bibnamefont
  {Schmidutz}}, \bibinfo {author} {\bibfnamefont {I.}~\bibnamefont
  {Gotlibovych}}, \bibinfo {author} {\bibfnamefont {R.~P.}\ \bibnamefont
  {Smith}}, \ and\ \bibinfo {author} {\bibfnamefont {Z.}~\bibnamefont
  {Hadzibabic}},\ }\bibfield  {title} {\enquote {\bibinfo {title}
  {{Bose-Einstein Condensation of Atoms in a Uniform Potential}},}\ }\href
  {\doibase 10.1103/PhysRevLett.110.200406} {\bibfield  {journal} {\bibinfo
  {journal} {Phys. Rev. Lett.}\ }\textbf {\bibinfo {volume} {110}},\ \bibinfo
  {pages} {200406} (\bibinfo {year} {2013})}\BibitemShut {NoStop}%
\bibitem [{\citenamefont {Chin}\ \emph {et~al.}(2010)\citenamefont {Chin},
  \citenamefont {Grimm}, \citenamefont {Julienne},\ and\ \citenamefont
  {Tiesinga}}]{Chin:2010}%
  \BibitemOpen
  \bibfield  {author} {\bibinfo {author} {\bibfnamefont {C.}~\bibnamefont
  {Chin}}, \bibinfo {author} {\bibfnamefont {R.}~\bibnamefont {Grimm}},
  \bibinfo {author} {\bibfnamefont {P.}~\bibnamefont {Julienne}}, \ and\
  \bibinfo {author} {\bibfnamefont {E.}~\bibnamefont {Tiesinga}},\ }\bibfield
  {title} {\enquote {\bibinfo {title} {Feshbach resonances in ultracold
  gases},}\ }\href {\doibase 10.1103/RevModPhys.82.1225} {\bibfield  {journal}
  {\bibinfo  {journal} {Rev. Mod. Phys.}\ }\textbf {\bibinfo {volume} {82}},\
  \bibinfo {pages} {1225--1286} (\bibinfo {year} {2010})}\BibitemShut {NoStop}%
\bibitem [{Note1()}]{Note1}%
  \BibitemOpen
  \bibinfo {note} {Due to geometric effects, in a harmonic trap the thermal
  atom number $N_{\protect \rm th}$ is not saturated even for very weak
  interactions~\cite {Tammuz:2011}, whereas in a box-trapped gas it is~\cite
  {Schmidutz:2014}.}\BibitemShut {Stop}%
  \bibitem [{\citenamefont {Tammuz}\ \emph {et~al.}(2011)\citenamefont {Tammuz},
  \citenamefont {Smith}, \citenamefont {Campbell}, \citenamefont {Beattie},
  \citenamefont {Moulder}, \citenamefont {Dalibard},\ and\ \citenamefont
  {Hadzibabic}}]{Tammuz:2011}   \BibitemOpen
  \bibfield  {author} {\bibinfo {author} {\bibfnamefont {N.}~\bibnamefont
  {Tammuz}}, \bibinfo {author} {\bibfnamefont {R.~P.}\ \bibnamefont {Smith}},
  \bibinfo {author} {\bibfnamefont {R.~L.~D.}\ \bibnamefont {Campbell}},
  \bibinfo {author} {\bibfnamefont {S.}~\bibnamefont {Beattie}}, \bibinfo
  {author} {\bibfnamefont {S.}~\bibnamefont {Moulder}}, \bibinfo {author}
  {\bibfnamefont {J.}~\bibnamefont {Dalibard}}, \ and\ \bibinfo {author}
  {\bibfnamefont {Z.}~\bibnamefont {Hadzibabic}},\ }\bibfield  {title}
  {\enquote {\bibinfo {title} {{Can a {B}ose Gas Be Saturated?}}}\ }\href
  {\doibase 10.1103/PhysRevLett.106.230401} {\bibfield  {journal} {\bibinfo
  {journal} {Phys. Rev. Lett.}\ }\textbf {\bibinfo {volume} {106}},\ \bibinfo
  {pages} {230401} (\bibinfo {year} {2011})}  \BibitemShut {NoStop}
  \bibitem [{\citenamefont {Weber}\ \emph {et~al.}(2003)\citenamefont {Weber},
  \citenamefont {Herbig}, \citenamefont {Mark}, \citenamefont {N\"agerl},\ and\
  \citenamefont {Grimm}}]{Weber:2003}%
  \BibitemOpen
  \bibfield  {author} {\bibinfo {author} {\bibfnamefont {T.}~\bibnamefont
  {Weber}}, \bibinfo {author} {\bibfnamefont {J.}~\bibnamefont {Herbig}},
  \bibinfo {author} {\bibfnamefont {M.}~\bibnamefont {Mark}}, \bibinfo {author}
  {\bibfnamefont {H.-C.}\ \bibnamefont {N\"agerl}}, \ and\ \bibinfo {author}
  {\bibfnamefont {R.}~\bibnamefont {Grimm}},\ }\bibfield  {title} {\enquote
  {\bibinfo {title} {{Three--Body Recombination at Large Scattering Lengths in
  an Ultracold Atomic Gas}},}\ }\href {\doibase 10.1103/PhysRevLett.91.123201}
  {\bibfield  {journal} {\bibinfo  {journal} {Phys. Rev. Lett.}\ }\textbf
  {\bibinfo {volume} {91}},\ \bibinfo {pages} {123201} (\bibinfo {year}
  {2003})}\BibitemShut {NoStop}%
\bibitem [{\citenamefont {Lewandowski}\ \emph {et~al.}(2003)\citenamefont
  {Lewandowski}, \citenamefont {McGuirk}, \citenamefont {Harber},\ and\
  \citenamefont {Cornell}}]{Lewandowski:2003}%
  \BibitemOpen
  \bibfield  {author} {\bibinfo {author} {\bibfnamefont {H.~J.}\ \bibnamefont
  {Lewandowski}}, \bibinfo {author} {\bibfnamefont {J.~M.}\ \bibnamefont
  {McGuirk}}, \bibinfo {author} {\bibfnamefont {D.~M.}\ \bibnamefont {Harber}},
  \ and\ \bibinfo {author} {\bibfnamefont {E.~A.}\ \bibnamefont {Cornell}},\
  }\bibfield  {title} {\enquote {\bibinfo {title} {{Decoherence--Driven Cooling
  of a Degenerate Spinor {B}ose Gas}},}\ }\href
  {https://link.aps.org/doi/10.1103/PhysRevLett.91.240404} {\bibfield
  {journal} {\bibinfo  {journal} {Phys. Rev. Lett.}\ }\textbf {\bibinfo
  {volume} {91}},\ \bibinfo {pages} {240404} (\bibinfo {year}
  {2003})}\BibitemShut {NoStop}%
\bibitem [{\citenamefont {Olf}\ \emph {et~al.}(2015)\citenamefont {Olf},
  \citenamefont {Fang}, \citenamefont {Marti}, \citenamefont {MacRae},\ and\
  \citenamefont {Stamper-Kurn}}]{Olf:2015}%
  \BibitemOpen
  \bibfield  {author} {\bibinfo {author} {\bibfnamefont {R.}~\bibnamefont
  {Olf}}, \bibinfo {author} {\bibfnamefont {F.}~\bibnamefont {Fang}}, \bibinfo
  {author} {\bibfnamefont {G.~E.}\ \bibnamefont {Marti}}, \bibinfo {author}
  {\bibfnamefont {A.}~\bibnamefont {MacRae}}, \ and\ \bibinfo {author}
  {\bibfnamefont {D.~M.}\ \bibnamefont {Stamper-Kurn}},\ }\bibfield  {title}
  {\enquote {\bibinfo {title} {{Thermometry and cooling of a {B}ose gas to 0.02
  times the condensation temperature}},}\ }\href {\doibase 10.1038/nphys3408}
  {\bibfield  {journal} {\bibinfo  {journal} {Nat. Phys.}\ }\textbf {\bibinfo
  {volume} {11}},\ \bibinfo {pages} {720} (\bibinfo {year} {2015})}\BibitemShut
  {NoStop}%
\bibitem [{Note2()}]{Note2}%
  \BibitemOpen
  \bibinfo {note} {One can repeat an analogous calculation for two-body losses
  due to, {\protect \it e.g.}, spin-changing collisions. In that case $\protect
  \mathaccentV {dot}05F{n}/n = -g_2 K_2 n$, with $g_2 = (n_0^2 + 4 n_0
  n_{\protect \rm th} + 2 n_{\protect \rm th}^2)/n^2 = 2-\eta ^2$. This gives
  $\protect \mathcal {P} = 6/(10-5\eta ^2)$, which can also be larger than
  1.}\BibitemShut {Stop}%
\bibitem [{\citenamefont {Smith}(2017)}]{Smith:2017}%
  \BibitemOpen
  \bibfield  {author} {\bibinfo {author} {\bibfnamefont {R.~P.}\ \bibnamefont
  {Smith}},\ }\enquote {\bibinfo {title} {Effects of {I}nteractions on
  {B}ose--{E}instein {C}ondensation},}\ in\ \href
  {https://doi.org/10.1017/9781316084366.008} {\emph {\bibinfo {booktitle}
  {Universal Themes of Bose--Einstein Condensation}}},\ \bibinfo {editor}
  {edited by\ \bibinfo {editor} {\bibfnamefont {N.~P.}\ \bibnamefont
  {Proukakis}}, \bibinfo {editor} {\bibfnamefont {D.~W.}\ \bibnamefont
  {Snoke}}, \ and\ \bibinfo {editor} {\bibfnamefont {P.~B.}\ \bibnamefont
  {Littlewood}}}\ (\bibinfo  {publisher} {Cambridge University Press},\
  \bibinfo {year} {2017})\ pp.\ \bibinfo {pages} {99--116}\BibitemShut
  {NoStop}%
\bibitem [{Note3()}]{Note3}%
  \BibitemOpen
  \bibinfo {note} {One can obtain identical results by iteratively removing, in
  small steps, particles according to Eq.~(\ref {eq_LossRates}) and energy
  according to Eq.~(\ref {eq_dEdt}), and then solving for the new equilibrium
  state under the constraints of the new total $N$ and $E$.}\BibitemShut
  {Stop}%
\bibitem [{\citenamefont {Shotan}\ \emph {et~al.}(2014)\citenamefont {Shotan},
  \citenamefont {Machtey}, \citenamefont {Kokkelmans},\ and\ \citenamefont
  {Khaykovich}}]{Shotan:2014}%
  \BibitemOpen
  \bibfield  {author} {\bibinfo {author} {\bibfnamefont {Z.}~\bibnamefont
  {Shotan}}, \bibinfo {author} {\bibfnamefont {O.}~\bibnamefont {Machtey}},
  \bibinfo {author} {\bibfnamefont {S.}~\bibnamefont {Kokkelmans}}, \ and\
  \bibinfo {author} {\bibfnamefont {L.}~\bibnamefont {Khaykovich}},\ }\bibfield
   {title} {\enquote {\bibinfo {title} {{T}hree--{B}ody {R}ecombination at
  {V}anishing {S}cattering {L}engths in an {U}ltracold {B}ose {G}as},}\ }\href
  {\doibase 10.1103/PhysRevLett.113.053202} {\bibfield  {journal} {\bibinfo
  {journal} {Phys. Rev. Lett.}\ }\textbf {\bibinfo {volume} {113}},\ \bibinfo
  {pages} {053202} (\bibinfo {year} {2014})}\BibitemShut {NoStop}%
\bibitem [{\citenamefont {Fattori}\ \emph {et~al.}(2008)\citenamefont
  {Fattori}, \citenamefont {D'Errico}, \citenamefont {Roati}, \citenamefont
  {Zaccanti}, \citenamefont {Jona-Lasinio}, \citenamefont {Modugno},
  \citenamefont {Inguscio},\ and\ \citenamefont {Modugno}}]{Fattori:2008}%
  \BibitemOpen
  \bibfield  {author} {\bibinfo {author} {\bibfnamefont {M.}~\bibnamefont
  {Fattori}}, \bibinfo {author} {\bibfnamefont {C.}~\bibnamefont {D'Errico}},
  \bibinfo {author} {\bibfnamefont {G.}~\bibnamefont {Roati}}, \bibinfo
  {author} {\bibfnamefont {M.}~\bibnamefont {Zaccanti}}, \bibinfo {author}
  {\bibfnamefont {M.}~\bibnamefont {Jona-Lasinio}}, \bibinfo {author}
  {\bibfnamefont {M.}~\bibnamefont {Modugno}}, \bibinfo {author} {\bibfnamefont
  {M.}~\bibnamefont {Inguscio}}, \ and\ \bibinfo {author} {\bibfnamefont
  {G.}~\bibnamefont {Modugno}},\ }\bibfield  {title} {\enquote {\bibinfo
  {title} {{A}tom {I}nterferometry with a {W}eakly {I}nteracting
  {B}ose--{E}instein {C}ondensate},}\ }\href
  {http://link.aps.org/doi/10.1103/PhysRevLett.100.080405} {\bibfield
  {journal} {\bibinfo  {journal} {Phys. Rev. Lett.}\ }\textbf {\bibinfo
  {volume} {100}},\ \bibinfo {pages} {080405} (\bibinfo {year}
  {2008})}\BibitemShut {NoStop}%
\bibitem [{\citenamefont {Monroe}\ \emph {et~al.}(1993)\citenamefont {Monroe},
  \citenamefont {Cornell}, \citenamefont {Sackett}, \citenamefont {Myatt},\
  and\ \citenamefont {Wieman}}]{Monroe:1993}%
  \BibitemOpen
  \bibfield  {author} {\bibinfo {author} {\bibfnamefont {C.~R.}\ \bibnamefont
  {Monroe}}, \bibinfo {author} {\bibfnamefont {E.~A.}\ \bibnamefont {Cornell}},
  \bibinfo {author} {\bibfnamefont {C.~A.}\ \bibnamefont {Sackett}}, \bibinfo
  {author} {\bibfnamefont {C.~J.}\ \bibnamefont {Myatt}}, \ and\ \bibinfo
  {author} {\bibfnamefont {C.~E.}\ \bibnamefont {Wieman}},\ }\bibfield  {title}
  {\enquote {\bibinfo {title} {{M}easurement of {C}s--{C}s elastic scattering
  at {T}=30 \ensuremath{\mu}{K}},}\ }\href {\doibase
  10.1103/PhysRevLett.70.414} {\bibfield  {journal} {\bibinfo  {journal} {Phys.
  Rev. Lett.}\ }\textbf {\bibinfo {volume} {70}},\ \bibinfo {pages} {414--417}
  (\bibinfo {year} {1993})}\BibitemShut {NoStop}%
\bibitem [{\citenamefont {{Bouchoule}}\ \emph {et~al.}(2018)\citenamefont
  {{Bouchoule}}, \citenamefont {{Schemmer}},\ and\ \citenamefont
  {{Henkel}}}]{Bouchoule:2018}%
  \BibitemOpen
  \bibfield  {author} {\bibinfo {author} {\bibfnamefont {I.}~\bibnamefont
  {{Bouchoule}}}, \bibinfo {author} {\bibfnamefont {M.}~\bibnamefont
  {{Schemmer}}}, \ and\ \bibinfo {author} {\bibfnamefont {C.}~\bibnamefont
  {{Henkel}}},\ }\bibfield  {title} {\enquote {\bibinfo {title} {{Cooling
  phonon modes of a {B}ose condensate with uniform few body losses}},}\ }\href
  {\doibase 10.21468/SciPostPhys.5.5.043} {\bibfield  {journal} {\bibinfo
  {journal} {SciPost Phys.}\ }\textbf {\bibinfo {volume} {5}},\ \bibinfo {eid}
  {043} (\bibinfo {year} {2018})}\BibitemShut {NoStop}%
%
\end{thebibliography}
\end{document}